\documentclass[aps,twocolumn]{revtex4}
\usepackage{amsfonts}
\usepackage{amsmath}
\usepackage{amssymb}
\usepackage{graphicx}
\usepackage{dcolumn}
\usepackage[sort&compress]{natbib}
\usepackage{subfigure}
\usepackage{ifpdf}
\usepackage{bm}
\usepackage{booktabs}
\usepackage{latexsym}
\usepackage{placeins}
\usepackage{longtable}

\begin{document}

\title{Generalized pricing formulas for stochastic volatility jump diffusion
models applied to the exponential Vasicek model}
\author{L. Z. Liang}
\affiliation{TQC, Universiteit Antwerpen, Universiteitsplein 1, 2610 Antwerpen, Belgium}
\author{D. Lemmens}
\affiliation{TQC, Universiteit Antwerpen, Universiteitsplein 1, 2610 Antwerpen, Belgium}
\author{J. Tempere}
\affiliation{TQC, Universiteit Antwerpen, Universiteitsplein 1, 2610 Antwerpen, Belgium}
\affiliation{Lyman Laboratory of Physics, Harvard University, Cambridge, MA 02138.}
\date{\today}

\begin{abstract}
Path integral techniques for the pricing of financial options are mostly based on models that can be recast in terms of a Fokker-Planck differential equation and that, consequently, neglect jumps and only describe drift and diffusion. We present a method to adapt formulas for both the path-integral propagators and the option prices themselves, so that jump processes are taken into account in conjunction with the usual drift and diffusion terms. In particular, we focus on stochastic volatility models, such as the exponential Vasicek model, and extend the pricing formulas and propagator of this model to incorporate jump diffusion with a given jump size distribution. This model is of importance to include non-Gaussian fluctuations beyond the Black-Scholes model, and moreover yields a lognormal distribution of the volatilities, in agreement with results from superstatistical analysis. The results obtained in the present formalism are checked with Monte Carlo simulations.
\end{abstract}

\maketitle

\section{Introduction}

\label{Introduction} It is well known that the pioneering option pricing
theory of Black and Scholes \cite{BS_1} and Merton \cite{BS_2} fails to
reflect some important empirical phenomena. Many studies have been conducted
to modify and improve the Black-Scholes model. Among others, popular models
include, (a) the local volatility models \cite{Derman_Kani}; (b) the
stochastic volatility (SV) models \cite{Hull_White, Stein, Heston}; (c) the
SV and stochastic interest rate models \cite{Amin_Ng, Bakshi_Chen, Scott,
Lemmens}; (d) the jump diffusion models \cite{Duffie_P_S, Merton, Kou}; (e) models based
on Levy process \cite{Geman, Schoutens, Carr, Wilmott, Cont}; and (f) the SV
jump diffusion models \cite{Bates, Bakshi_Cao_Chen, Andersen_Benzoni_Lund,
Pan, Eraker, Chernov, Sepp}.

Inspired by \cite{Andersen_Benzoni_Lund, Bates, Bakshi_Cao_Chen, Cont,
Gatheral} we will focus on the latter class of models.\ For example, Cont and
Tankov \cite{Cont} and Gatheral \cite{Gatheral} motivate that the
combination of jumps in returns and SV makes it possible to calibrate the
implied volatility surface, without using time dependent parameters. Jumps
make it possible to reproduce strong skews and smiles at short maturities
while SV provides for the calibration of the term structure, especially for
long-term smiles.

In this article we will present a method that makes it possible to extend
the Fourier space propagator of a
general SV model to the Fourier space propagator of that SV model where an arbitrary jump
process has been added to the asset price dynamics. Thereby we contribute to
the existing work on Fourier transform methods applied to option pricing.
For example in \cite{Duffie_P_S} jump diffusions are treated and prices for
some exotic options are obtained. In \cite{Yan}\ the Heston model is
extended with a jump process for the asset price. In \cite{Sepp}\ the Heston
model is extended with arbitrary jump processes in both the asset price\ and
the volatility process.

As an application, we investigate a model where we assume that the stochastic volatility follows an exponential Vasicek model
\cite{Liu, Micciche}. To the best of our knowledge, for this model no closed form formulas for the propagator or
the vanilla option price\ exist yet. Making use of path integral methods
\cite{Baaquie, Kleinert, Lemmens} we derive approximative closed form
formulas for the propagator and for vanilla option prices for this model (for more
information about methods from physics applied to finance see for example \cite%
{Dash, Voit, Straeten}). Using Monte Carlo (MC) simulations we specify
parameter ranges for which the approximation is valid. Using the above
mentioned method we extend the propagator of this model to the propagator of this
model extended with jumps in the asset price which leads also to closed form
pricing formulas in this extended model. Also these last results are checked
with MC simulations.

This paper is organized as follows. In section \ref{s.Results} we present
the method for extending the propagator of a general SV model to the propagator of that
model with jumps in the asset price.  In section \ref{s.LN}, we present an approximative propagator for
jump diffusion models where the volatility is assumed to follow an
exponential Vasicek model. Section \ref{s.European Option Pricing} is
devoted to European vanilla option pricing, as well as comparisons with MC
simulations. In this section we also give parameter ranges for the
approximation made in the exponential Vasicek model to be valid. And finally
a conclusion is given in section \ref{s.Conclusion}.

\section{General Propagator Formulas}\label{s.Results}

\subsection{Arbitrary SV models}

We assume that the asset price process $S(t)$ follows the Black-Scholes
stochastic differential equation (SDE):
\begin{equation} \label{e.ds}
dS(t)=rS(t)dt+\sigma (t)S(t)dW_{1}(t),
\end{equation}%
in which $r$ is the constant interest rate and the volatility $\sigma (t)$
is behaving stochastically over time, following an arbitrary stochastic
process:
\begin{equation}
d\sigma (t)=A(t,\sigma (t))dt+B(t,\sigma (t))dW_{2}(t).  \label{e.dsgn}
\end{equation}%

Here and in the rest of the article $W_{j}=\{W_{j}(t),t\geq 0\} \,(j=1,2)$
are two correlated Wiener processes such that Cov$\left[ \,dW_{1}(t)%
\,dW_{2}(t)\, \right] =\rho \,dt$.

Eq.(\ref{e.ds}) is commonly expressed as a function of the logreturn $x(t)=\ln S(t),$ which leads to a new SDE:
\begin{equation}
dx(t)=\left( r-\frac{1}{2}\sigma ^{2}(t)\right) dt+\sigma (t)dW_{1}(t).
\end{equation}

To deal with the pricing problem, we need to solve for the propagator of the
joint dynamics of $x(t)$ and $\sigma (t)$. The propagator, denoted by $%
\mathcal{P}(x_{T},\sigma _{T},T|\,x_{0},\sigma _{0},0)$, describes the
probability that $x$ has the value $x_{T}$ and $\sigma $ has the value $%
\sigma _{T}$ at a later time $T$ given the initial values $x_{0}$ and $%
\sigma _{0}$ respectively at time $0$. It satisfies the following
Kolmogoroff forward equation:
\begin{eqnarray}\label{e.FW}
\frac{\partial \mathcal{P}}{\partial T} &=&\frac{\partial }{\partial x_{T}}\left[ -(r-\frac{1}{2}\sigma _{T}^{2})%
\mathcal{P}\right] +\frac{1}{2}\frac{\partial ^{2}}{\partial x_{T}^{2}}%
\left[ \sigma _{T}^{2}\mathcal{P}\right]  \notag \\
&&+\frac{\partial }{\partial \sigma _{T}}\left[ -A(T,\sigma _{T})\mathcal{P}%
\right] +\frac{1}{2}\frac{\partial ^{2}}{\partial \sigma _{T}^{2}}\left[
B^{2}(T,\sigma _{T})\mathcal{P}\right]  \notag \\
&&+\rho \frac{\partial ^{2}}{\partial x_{T}\, \partial \sigma _{T}}\left[
\sigma _{T}B(T,\sigma _{T})\mathcal{P}\right] ,
\end{eqnarray}%
with initial condition
\begin{equation}
\mathcal{P}(x_{T},\sigma _{T},0|\,x_{0},\sigma _{0},0)=\delta
(x_{T}-x_{0})\, \delta (\sigma _{T}-\sigma _{0}).
\end{equation}

\subsection{SV jump diffusion models}

A general SV jump diffusion model is obtained by adding an arbitrary jump
process into the asset price process (see for instance \cite{Bates}). That is, equation (\ref{e.ds}) becomes
\begin{equation}
dS(t)=\mu S(t)dt+\sigma (t)S(t)dW_{1}(t)+\left( e^{J}-1\right) S(t)dN(t),
\end{equation}%
where $N=\{N(t),t\geq 0\}$ is an independent Poisson process with intensity
parameter $\lambda >0$, i.e. $\mathbb{E}[\,N(t)\,]=\lambda \,t$. The random
variable $J$ with probability density $\varpi (J)$ describes the magnitude
of the jump when it occurs.

Here the risk-neutral drift $\mu =r-\lambda \,m^{j}$ is no longer the
constant interest rate $r$, rather it is adjusted by a compensator term $%
\lambda \,m^{j}$, with $m^{j}$ the expectation value of $e^{J}-1$:
\begin{equation}
m^{j}=\mathbb{E}\left[ e^{J}-1\right] =\int_{-\infty }^{+\infty
}(e^{J}-1)\varpi (J)dJ,
\end{equation}%
so that the asset price process constitutes a martingale under the risk
neutral measure. And the logreturn $x(t)$ follows a new SDE:
\begin{equation}
    dx(t) = \left(r - \lambda m^j - \frac{1}{2} \sigma^2(t) \right) dt + \sigma(t) dW_1(t) + J dN(t).
\end{equation}

Given the same arbitrary SV process (\ref{e.dsgn}), the new propagator of this model, denoted by $\mathcal{P}%
_{J}(x_{T},\sigma _{T},T|x_{0},\sigma _{0},0)$, satisfies the new
Kolmogoroff forward equation (see for instance \cite{Gardiner})
\begin{eqnarray}
\frac{\partial \mathcal{P}_{J}}{\partial T} & = &\frac{\partial }{\partial x_{T}}\left[ -\left( r-\lambda m^{j}-\frac{1}{2}%
\sigma _{T}^{2}\right) \mathcal{P}_{J}\right]  \notag \\
&&+\frac{1}{2}\frac{\partial ^{2}}{\partial x_{T}^{2}}\left[ \sigma _{T}^{2}%
\mathcal{P}_{J}\right] +\frac{\partial }{\partial \sigma _{T}}\left[
-A(T,\sigma _{T})\mathcal{P}_{J}\right]  \notag \\
&&+\frac{1}{2}\frac{\partial ^{2}}{\partial \sigma _{T}^{2}}\left[
B^{2}(T,\sigma _{T})\mathcal{P}_{J}\right]  \notag \\
&&+\rho \frac{\partial ^{2}}{\partial x_{T}\, \partial \sigma _{T}}\left[
\sigma _{T}B(T,\sigma _{T})\mathcal{P}_{J}\right]  \notag \\
&&+\lambda \int_{-\infty }^{+\infty }\left[ \mathcal{P}_{J}(x_{T}-J)-%
\mathcal{P}_{J}(x_{T})\right] \varpi (J)dJ.  \label{fkarj6}
\end{eqnarray}

If we write the propagator of the arbitrary SV model as a Fourier integral
(here and below, $i$ is the imaginary unit)
\begin{eqnarray}
&&\mathcal{P}(x_{T},\sigma _{T},T|x_{0},\sigma _{0},0)  \notag
\label{e.pAR} \\
&=&\int_{-\infty }^{+\infty }\frac{dp}{2\pi }e^{ip(x_{T}-x_{0})}F(%
\sigma _{T},\sigma _{0}, r, p, T),
\end{eqnarray}%
then the propagator of arbitrary SV jump diffusion models can be written as
\begin{eqnarray}
&&\mathcal{P}_{J}(x_{T},\sigma _{T},T|x_{0},\sigma _{0},0)  \notag \\
&=&\int_{-\infty }^{+\infty }\frac{dp}{2\pi }e^{ip(x_{T}-x_{0})}F(%
\sigma _{T},\sigma _{0}, r, p, T)\,e^{U(p,T)},  \label{algeres}
\end{eqnarray}%
where
\begin{equation}
U(p,T)=\lambda T\int_{-\infty }^{+\infty }\left[ e^{-ipJ}-1+ip\left(
e^{J}-1\right) \right] \varpi (J)dJ.  \label{e.U}
\end{equation}%
The proof of this statement is given in the Appendix \ref{Appendx}. Note the
relation between propagators (\ref{e.pAR}) and (\ref{algeres}). The only
difference between them is the factor $e^{U(p,T)}$.

If this is applied to the propagator of the Heston model
\cite{Lemmens}, the propagator of the Heston model with jumps is obtained. This
propagator is similar as the one derived in Ref. \cite{Sepp}.
Furthermore the above described method can be combined with the method
described in Ref. \cite{Lemmens}\ for finding the propagator of a model including both
SV and stochastic interest rate. In particular extending the result of Ref. \cite
{Lemmens} for the Heston model with stochastic interest rate to include
jumps again only involves multiplying the propagator with\ $e^{U(p,T)}$\ as in (%
\ref{algeres}). In the next section, as an example of the method of this
section the volatility of the asset price will be assumed to follow an
exponential Vasicek model.

\section{Exponential Vasicek SV model with price jumps}\label{s.LN}

The Heston model assumes that the squared volatility follows a CIR process
which has a gamma distribution as stationary distribution. This assumption should be
compared with market data. Attempts to reconstruct the stationary
probability distribution of volatility from the time series data (among
others, see Refs. \cite{Liu, Micciche, Straeten}) generally agree that the central
part of the stationary volatility distribution is better described by a
lognormal distribution.

Due to the different structure in path-behavior between different models,
Schoutens, Simons and Tistaert find that the resulting exotic prices can
vary significantly \cite{Wilmott}. So an investigation into an alternative
model which fits market data better is meaningful.

Furthermore the model will serve here both to demonstrate the use of path
integral methods in finance and to illustrate the method of section \ref%
{s.Results}.

When $\sigma (t)$\ is assumed to be an exponential Vasicek process (used for
example by Chesney and Scott \cite{Chesney_Scott}), this results in the
following two SDEs
\begin{eqnarray}
dS &=&rS\,dt+\sigma \,S\,dW_{1}, \\
d\sigma  &=&\sigma \left( \beta \left[ \bar{a}-\ln \sigma \right] +\frac{1}{2%
}\gamma ^{2}\right) dt+\gamma \sigma dW_{2}.
\end{eqnarray}%
This model has a lognormal stationary volatility distribution and we will
denote it by the LN model, the propagator for this model will be denoted by
$\mathcal{P}_{LN}$. In this model $\ln \sigma (t)$ is a mean
reverting process, with $\beta $ the spring constant of the force that
attracts the logarithm of asset volatility to its mean reversion level $\bar{%
a}$. Again $\gamma $ is the volatility of the asset volatility. As far as we
know, there is no closed form option pricing formula for this model. In this
section, we will give an approximation for the propagator of this model. In the
next section we will give an approximation for the vanilla option price and determine
a parameter range for which the approximation is good. The derivation starts
with the following substitutions:
\begin{eqnarray}
y(t) &=&x(t)-\frac{\rho }{\gamma }\,e^{z(t)}-rt,  \\
z(t) &=&\ln \sigma (t),
\end{eqnarray}%
where $x(t)$\ is defined as before. This leads to two uncorrelated
equations:
\begin{eqnarray}
dy &=&\left[ -\frac{1}{2}\,e^{2z}-\rho \left( \frac{\beta (\bar{a}-z)}{%
\gamma }+\frac{\gamma }{2}\right) e^{z}\right] dt  \notag \\
&&+e^{z}\sqrt{1-\rho ^{2}}dB_{1},  \\
dz &=&\beta \left( \bar{a}-z\right) dt+\gamma dB_{2},  \label{e.dzeta}
\end{eqnarray}%
where $B_{1}$ and $B_{2}$\ are two uncorrelated Wiener processes. Since
these equations are uncorrelated, the propagator $\mathcal{P}%
_{LN}(y_{T},z_{T}|\,y_{0},z_{0})$ is given by the following path integral
\begin{eqnarray}
&&\mathcal{P}_{LN}(y_{T},z_{T}|\,y_{0},z_{0})  \notag \\
&=&\int \mathcal{D}z\, \left( \int \mathcal{D}y e^{-\int_{0}^{T}\mathcal{L}%
[y,z]dt}\right) e^{-\int_{0}^{T}\mathcal{L}[z]dt},  \label{LN_Propagator}
\end{eqnarray}%
where the Lagrangians are given by:
\begin{eqnarray}
\mathcal{L}[y,z] &=&\frac{\left[ \dot{y}+\frac{1}{2}\,e^{2z}+\rho \left(
\frac{\beta (\bar{a}-z)}{\gamma }+\frac{\gamma }{2}\right) e^{z}\right] ^{2}%
}{2(1-\rho ^{2})\,e^{2z}},  \\
\mathcal{L}[z] &=&\frac{\left[ \dot{z}-\beta \left( \bar{a}-z\right) \right]
^{2}}{2\gamma ^{2}}-\frac{\beta }{2}.
\end{eqnarray}%
The first step in the evaluation of (\ref{LN_Propagator}) is the integration
over all $y$ paths. Because the action is quadratic in $y$, this path
integration can be done analytically and yields
\begin{eqnarray}
&&\mathcal{P}_{LN}(y_{T},z_{T}|\,y_{0},z_{0})  \notag \\
&=&\int \mathcal{D}z\,e^{-\int_{0}^{T}\mathcal{L}[z]dt}\, \frac{1}{\sqrt{2\pi
(1-\rho ^{2})\int_{0}^{T}e^{2z}dt}}  \notag \\
&&\times \,e^{-\frac{\left[ y_{T}-y_{0}+\frac{1}{2}\int_{0}^{T}e^{2z}dt+\rho
\int_{0}^{T}\left( \frac{\beta (\bar{a}-z)}{\gamma }+\frac{\gamma }{2}%
\right) e^{z}dt\right] ^{2}}{2(1-\rho ^{2})\int_{0}^{T}e^{2z}dt}}.
\end{eqnarray}%
Note that the probability to arrive in $(y_{T},z_{T})$ only depends on the
average value of the volatility along the path $z(t)$, in agreement with
Ref. \cite{Chesney_Scott}. With the help of a Fourier transform, we rewrite
the preceding expression as follows
\begin{eqnarray}
&&\mathcal{P}_{LN}(y_{T},z_{T}|\,y_{0},z_{0})  \notag \\
&=&\int_{-\infty }^{+\infty }\frac{dp}{2\pi }\,e^{ip(y_{T}-y_{0})}\int
\mathcal{D}z \,e^{-\int_{0}^{T}\mathcal{L}[z]dt}  \notag \\
&&\times \,e^{-\frac{(1-\rho ^{2})p^{2}-ip}{2}\int_{0}^{T}e^{2z}dt+ip\rho
\int_{0}^{T}\left( \frac{\beta (\bar{a}-z)}{\gamma }+\frac{\gamma }{2}%
\right) e^{z}dt}.   \label{onbdmpi}
\end{eqnarray}

If $\zeta (t)=z(t)-\bar{a},$\ then $\zeta (t)$\ is close to zero because $%
z(t)$ is a mean reverting process with mean reversion level $\bar{a}$, This
motivates the approximation $e^{\zeta }\approx 1+\zeta +\frac{\zeta ^{2}}{2}$.
This type of approximation is akin to expanding the path integral around the saddle point up to second order in the fluctuations, as in the Nozieres-Schmitt-Rink formalism \cite{Nozieres} extended to path-integration by Sa de Melo, Randeria and Engelbrecht \cite{Sa}.
Now we can work out the remaining path integral in (\ref{onbdmpi})%
\begin{eqnarray}
&&\int \mathcal{D}z\,e^{-\int_{0}^{T}\left[ \mathcal{L}[z]+\frac{(1-\rho
^{2})p^{2}-ip}{2}e^{2z}-ip\rho \left( \frac{\beta (\bar{a}-z)}{\gamma }+%
\frac{\gamma }{2}\right) \,e^{z}\right] dt}  \notag \\
&=&\int \mathcal{D}\zeta \,e^{-\int_{0}^{T}\left \{ \frac{\left[ \dot{\zeta%
}+\beta \zeta \right] ^{2}}{2\gamma ^{2}}-\frac{\beta }{2}+\frac{A}{2}%
\,e^{2\zeta }+B\beta \zeta \,e^{\zeta }-\frac{B\gamma ^{2}}{2}\,e^{\zeta
}\right \} dt}  \notag \\
&=&e^{\frac{\omega \left[ \left( \zeta _{T}+\frac{\gamma ^{2} M }{\omega
^{2}}\right) ^{2}-\left( \zeta _{0}+\frac{\gamma ^{2} M }{\omega ^{2}}%
\right) ^{2}\right] -\beta \left( \zeta _{T}^{2}-\zeta _{0}^{2}\right) }{%
2\gamma ^{2}}}  \notag \\
&&\times e^{\left[ \frac{\beta -\omega -A+B\gamma ^{2}}{2}+\frac{\gamma
^{2} M^{2}}{2\omega ^{2}}\right] T}  \notag \\
&&\times \sqrt{\frac{\omega }{\pi \gamma ^{2}(1-e^{-2\omega T})}}\,e^{-\frac{%
\omega \left[ \left( \zeta _{T}+\frac{\gamma ^{2}M }{\omega ^{2}}%
\right) -\left( \zeta _{0}+\frac{\gamma ^{2} M}{\omega ^{2}}\right)
\,e^{-\omega T}\right] ^{2}}{\gamma ^{2}(1-e^{-2\omega T})}},
\end{eqnarray}%
where
\begin{eqnarray}
A &=&\left[ (1-\rho ^{2})p^{2}-ip\right] e^{2\bar{a}}, \\
B &=&ip\rho \frac{1}{\gamma }\,e^{\bar{a}}, \\
\omega  &=&\sqrt{\beta ^{2}+2\gamma ^{2}\left( A+B\beta -\frac{B\gamma ^{2}}{%
4}\right) },  \\
M  &=&A+B\beta -\frac{B\gamma ^{2}}{2}.
\end{eqnarray}

We see that also the integral over the final value $\zeta _{T}$ can be done,
yielding the marginal probability distribution:
\begin{eqnarray}
&&\mathcal{P}_{LN}(x_{T}|x_{0},\zeta _{0})  \notag  \label{P_LN} \\
&=&\int_{-\infty }^{+\infty }\frac{dp}{2\pi }e^{ip\left[ x_{T}-x_{0}-rT%
\right] +B\left( e^{\zeta _{0}}-1\right) } \notag \\
&&\times e^{\frac{\beta \zeta _{0}^{2}-\omega \left( \zeta _{0}+\frac{\gamma
^{2} M }{\omega ^{2}}\right) ^{2}}{2\gamma ^{2}}+\left[ \frac{\beta
-\omega -A+B\gamma ^{2}}{2}+\frac{\gamma ^{2}  M^{2}}{2\omega ^{2}}%
\right] T}  \notag \\
&&\times \frac{e^{\frac{\Xi }{\gamma ^{2}\left[ 2\omega +\left( \beta
-\omega +B\gamma ^{2}\right) (1-e^{-2\omega T})\right] }}}{\sqrt{1+\frac{%
1-e^{-2\omega T}}{2\omega }\left[ \beta -\omega +B\gamma ^{2}\right] }},
\end{eqnarray}%
where
\begin{eqnarray}
\Xi  &=&\omega \left[ 2B\gamma ^{2} N+\omega (N-\frac{\gamma ^{2} M }{%
\omega ^{2}})^{2}-(\beta +B\gamma ^{2}) N^{2}\right]   \notag \\
&&+(1-e^{-2\omega T})\left[ \frac{B^{2}\gamma ^{4}}{2}-\frac{B\gamma
^{4} M }{\omega }\right.   \notag \\
&&\quad \quad \quad \quad \quad \quad \quad \left. +\frac{(\beta +B\gamma
^{2})\gamma ^{4} M^{2}}{2\omega ^{3}}\right] , \\
N &=&\frac{\gamma ^{2}  M}{\omega ^{2}}-(\zeta _{0}+\frac{\gamma
^{2} M }{\omega ^{2}})e^{-\omega T}.
\end{eqnarray}

\begin{figure}[t]
\includegraphics[width= 1\linewidth]{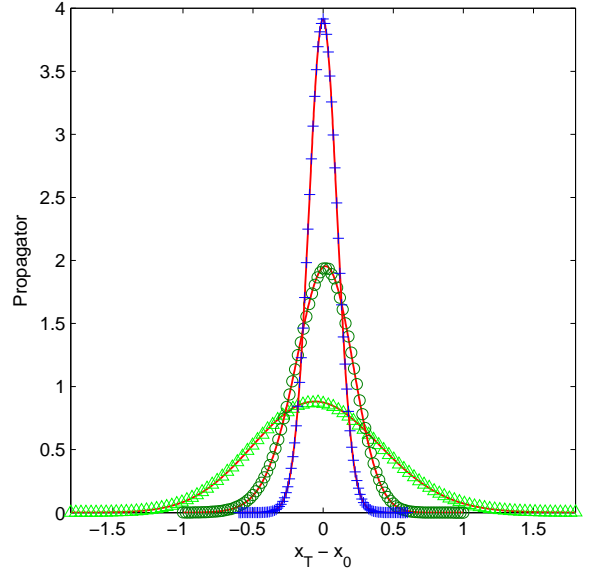}\newline
\caption{Propagator $\mathcal{P} (x_T | x_0, \protect \zeta_0)$ as a function
of $x_T - x_0$. The full curves are our analytical results, while the symbols represent Monte Carlo simulations. $T$ = 0.25y, $\protect \rho$ = 0 (crosses). $T$ =
1y, $\protect \rho$ = -0.5 (circles). $T$ = 5y, $\protect \rho$ = 0.5 (triangles).
For the other parameters the following values are used for the three figures: $\protect \beta = 5, \bar{a} = -1.6, \protect \gamma = 0.5, r = 0.015$.}
\label{f.MC_P_LN}
\end{figure}

The goodness of this approximative propagator needs the support from MC
simulations because of the lack of a closed form solution.

Figure \ref{f.MC_P_LN} shows the propagators as a function of $x_{T}-x_{0}$%
, i.e., $\ln \frac{S_{T}}{S_{0}}$. The full curves come from expression (\ref%
{P_LN}), while the marked ones are MC simulation results, with time to
maturity ranging from three months to five years, and correlation coefficients $%
0$, $-0.5$ and $0.5$ respectively. Here and in the rest of the article we
will set $\sigma _{0}$\ equal to the long time average of the volatility:%
\begin{equation}
\sigma _{0}=\lim \limits_{t\rightarrow \infty }\mathbb{E}\left[ \sigma (t)%
\right] =\exp \left \{ \bar{a}+\frac{\gamma ^{2}}{4\beta }\right \} ,
\end{equation}%
which seems a reasonable choice. For these MC simulations 5,000,000 sample
paths are used.

It is seen that our analytical results fit the MC simulations quite well.
Actually, using the parameters of Fig. \ref{f.MC_P_LN}, and putting expression (\ref%
{P_LN}) for those three cases into the left hand side of the Kolmogorov
backward equation:
\begin{eqnarray}
-\frac{\partial \mathcal{P}}{\partial T} &+&\left[ r-\frac{1}{2}\,e^{2(\zeta
_{0}+\bar{a})}\right] \frac{\partial \mathcal{P}}{\partial x_{0}}+\frac{1}{2}%
\,e^{2(\zeta _{0}+\bar{a})}\frac{\partial ^{2}\mathcal{P}}{\partial x_{0}^{2}%
}  \notag \\
-\beta \zeta _{0}\frac{\partial \mathcal{P}}{\partial \zeta _{0}} &+&\frac{1%
}{2}\gamma ^{2}\frac{\partial ^{2}\mathcal{P}}{\partial \zeta _{0}^{2}}+\rho
\,e^{\zeta _{0}+\bar{a}}\gamma \frac{\partial ^{2}\mathcal{P}}{\partial
x_{0}\zeta _{0}}=0,
\end{eqnarray}%
we find that, for different $x_{T}$ values, the absolute values are all in
the order of $10^{-7}$ or even smaller. In section \ref{s.mcsim} we come
back to the discussion concerning the goodness of our approximation.

According to the discussion of Section \ref{s.Results}, an extension of this
model to the one with price jumps is straightforward: the new marginal
probability distribution would be:
\begin{eqnarray}
&&\mathcal{P}_{LNJ}(x_{T}|0,\zeta _{0})  \notag \\
&=&\int_{-\infty }^{+\infty }\frac{dp}{2\pi }e^{ip\left[ x_{T}-x_{0}-rT%
\right] +B\left( e^{\zeta _{0}}-1\right) } \notag \\
&&\times e^{\frac{\beta \zeta _{0}^{2}-\omega \left( \zeta _{0}+\frac{\gamma
^{2} M }{\omega ^{2}}\right) ^{2}}{2\gamma ^{2}}+\left[ \frac{\beta
-\omega -A+B\gamma ^{2}}{2}+\frac{\gamma ^{2} M^{2}}{2\omega ^{2}}%
\right] T}  \notag \\
&&\times \frac{e^{\frac{\Xi }{\gamma ^{2}\left[ 2\omega +\left( \beta
-\omega +B\gamma ^{2}\right) (1-e^{-2\omega T})\right] }}}{\sqrt{1+\frac{%
1-e^{-2\omega T}}{2\omega }\left[ \beta -\omega +B\gamma ^{2}\right] }} \notag\\
&&\times e^{\lambda T\int_{-\infty }^{+\infty }\left[ e^{-ipJ}-1+ip\left(
e^{J}-1\right) \right] \varpi (J)dJ},
\end{eqnarray}
where the same notations as in Eq.(\ref{P_LN}) are used.

\section{European Vanilla Option Pricing}

\label{s.European Option Pricing}

\subsection{General Pricing Formulas}

If we denote the general marginal propagator by
\begin{equation}
\mathcal{P}(x_{T}|x_{0},\sigma _{0})=\int_{-\infty }^{+\infty }\frac{dp}{%
2\pi }\,e^{ip(x_{T}-x_{0}-rT)}F(p,T)\,e^{U(p,T)},
\end{equation}%
then the option pricing formula of a vanilla call option $\mathcal{C}$ with
expiration date $T$ and strike price $K$ is given by the discounted
expectation value of the payoff:
\begin{eqnarray}  \label{e.C}
\mathcal{C} &=&e^{-rT}\int_{-\infty }^{+\infty }\left( e^{x_{T}}-K\right)
_{+}\mathcal{P}(x_{T}|x_{0},\sigma _{0})dx_{T}  \notag  \label{C} \\
&=&\frac{\mathcal{G}(0)}{2}+i\int_{-\infty }^{+\infty }\frac{dp}{2\pi }\frac{%
e^{ip\left( \ln \frac{K}{S_{0}}-rT\right) }\mathcal{G}(p)}{p},
\end{eqnarray}%
where
\begin{eqnarray} \label{e.G}
\mathcal{G}(p) &=&S_{0}F(p+i,T)\,e^{U(p+i,T)}  \notag \\
&&-K\,e^{-rT}F(p,T)\,e^{U(p,T)}, \
\end{eqnarray}%
and $\left( x\right) _{+}=\max \left( x,0\right)$.

Here we have followed the derivation outlined in Ref. \cite{Kleinert2}. In
particular for the LN model $F(p,T)$ equals:
\begin{eqnarray}  \label{e.F}
F(p,T)&=& e^{\frac{\beta \zeta _{0}^{2}-\omega \left( \zeta _{0}+\frac{\gamma
^{2}  M }{\omega ^{2}}\right) ^{2}}{2\gamma ^{2}}+\left[ \frac{\beta
-\omega -A+B\gamma ^{2}}{2}+\frac{\gamma ^{2} M^{2}}{2\omega ^{2}}
\right] T}  \notag \\
&&\times \frac{e^{B (e^{\zeta_0} - 1) + \frac{\Xi }{\gamma ^{2}\left[ 2\omega +\left( \beta
-\omega +B\gamma ^{2}\right) (1-e^{-2\omega T})\right] }}}{\sqrt{1+\frac{%
1-e^{-2\omega T}}{2\omega }\left[ \beta -\omega +B\gamma ^{2}\right] }}.
\end{eqnarray}%

At this stage one needs to specify the PDF for the jump sizes. Merton \cite%
{Merton} and Kou \cite{Kou} proposed a normal distributed jump size, denoted
by $\varpi _{M}(J)$, and a asymmetric double exponential distributed one,
denoted by $\varpi _{K}(J)$, respectively:
\begin{eqnarray}
\varpi _{M}(J) &=&\frac{1}{\sqrt{2\pi \delta ^{2}}}\,e^{-\frac{(J-\nu )^{2}}{%
2\delta ^{2}}},  \\
\varpi _{K}(J) &=&p_{+}\frac{1}{\eta _{+}}\,e^{-\frac{1}{\eta _{+}}J}\Theta \left( J\right)   \notag \\
&&+\,p_{-}\frac{1}{\eta _{-}}\,e^{\frac{1}{\eta _{-}}J}\, \Theta%
\left( -J\right) .
\end{eqnarray}%
For the Merton model $\nu $\ is the mean jump size and $\delta $\ is the
standard deviation of the jump size. For Kou's model $0<\eta _{+}<1$, $\eta
_{-}>0$ are means of positive and negative jumps respectively.\ $p_{+}$\ and
$p_{-}$\ represent the probabilities of positive and negative jumps, $p_{+}>0
$, $p_{-}>0$, $p_{+}+$ $p_{-}=1$ and $\Theta $\ is the Heaviside function.

According to expression (\ref{e.U}), it is easy to derive their
corresponding $U(p,T)$'s:
\begin{eqnarray}
U_{M}(p,T) &=&\lambda T\left[ e^{-ip\nu -\frac{1}{2}\delta
^{2}p^{2}}-1\right.   \notag \\
&&\quad \quad \left. +ip\left( e^{\nu +\frac{1}{2}\delta ^{2}}-1\right) %
\right] ,  \label{e.UM}   \\
U_{K}(p,T) &=&\lambda T\left[ \frac{p_{+}}{1+ip\eta _{+}}+\frac{p_{-}}{%
1-ip\eta _{-}}-1\right.   \notag \\
&&\quad \quad \left. +ip(\frac{p_{+}}{1-\eta _{+}}+\frac{p_{-}}{1+\eta _{-}}%
-1)\right] .  \label{e.UK}
\end{eqnarray}

Using expression (\ref{e.F}) and results (\ref{e.UM}), (\ref{e.UK}) in formulas (\ref{e.C}), (\ref{e.G}) allows to find the price of the vanilla call option for the exponential Vasicek stochastic volatility with price jumps model.

\subsection{Monte Carlo simulations\label{s.mcsim}}

To test our analytical pricing formula for the LN model, we focus on the
parameters that most strongly influence the approximation. To satisfy the
assumption that quadratic fluctuations around the mean reversion level $\bar{a}$ captures the behavior of the volatility well, the mean
reversion speed $\beta $ and the volatility $\gamma $ of asset volatility
are crucial.

The substitution $\tau =\gamma ^{2}t$ transforms expression (\ref{e.dzeta})
into
\begin{equation}
dz(\tau )=\frac{\beta }{\gamma ^{2}}\left[ \bar{a}-z(\tau )\right] d\tau
+dB_{2}(\tau ),   \\
\end{equation}%
showing that it is actually the parameter $c=\frac{\beta }{\gamma ^{2}}$
which determines whether the approximation will be good. For bigger $c$\
values the approximation $z\left( t\right) \approx \bar{a}$ will be better.

As the correlation parameter $\rho $ controls the skewness of spot returns,
we will also consider the typical negative and positive skewed cases by
taking values $-0.5$, $0$\ and $0.5$ for this parameter. On the other hand,
the constant interest rate $r$ and the mean reversion level $\bar{a}$ do not
influence the accuracy of the result a lot, and we just assume them to be constant values: $r = 0.015$ and $\bar{a} = -1.6 \approx \ln 0.2$.
These two parameters seem to be quite reasonable for the present European options.

\begin{figure*}[ht]
\includegraphics[width=1\textwidth]{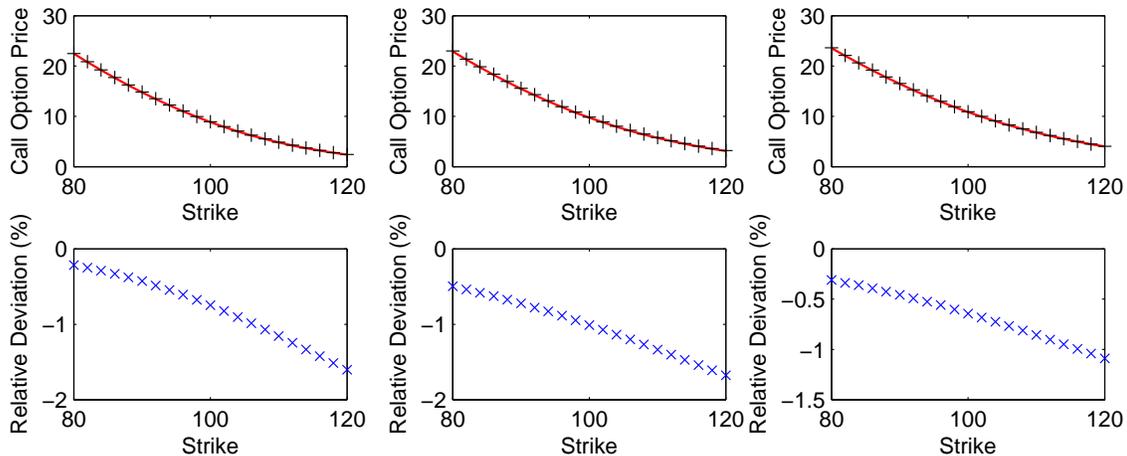}\newline
\caption{The upper figures show European call option prices in the LN model
(left), the LN model with Merton's jump (middle) and the LN model with Kou's
jump (right). The red curves are our analytical results and the black crosses
are the Monte Carlo simulations. The corresponding lower figures give the
relative deviations of our analytical results from the MC simulations in the
unit of percent. Parameter values $S_0 = 100$, $r = 0.015$, $T = 1$, $%
\protect \beta = 5$, $\bar{a} = -1.6$, $\protect \gamma = 0.5$, $\protect \rho %
= -0.5$, $\protect \lambda = 10$, $\protect \nu = -0.01$, $\protect \delta =
0.03$, $p_+ = 0.3$, $p_- = 0.7$, $\protect \eta_+ = 0.02$, $\protect \eta_- =
0.04$ are used here.}
\label{f.call}
\end{figure*}

To get an idea of what is a reasonable range for $c$,\ and since calibration
values for the LN model are not available, we took calibration values from
the literature \cite{Heston,Yacine} for the Heston model and fitted our
model to the volatility distribution of the Heston model with those
parameters. For \cite{Heston}\ we obtained $c\approx 7$ and for \cite{Yacine}
$c\approx 18$. Therefore in Table I we used values for $\beta $ and $\gamma $%
\ such that $c$\ ranges from $4.08$ up to $25$.\ We calculated prices for $%
S_{0}=100$\ and $K=90,$ $100$\ and $110$.

The comparison of our analytical solution with the MC solution for a
European call option in the LN model as shown in Table \ref{T_1} suggests
that for the above mentioned parameter values the relative errors are less
than $3\%$ and most of the time even less than $1\%$, which is acceptable
when we take the typical bid-ask spread for European options into account.
Here each MC simulation runs 20,000,000 times.

For the basic LN model we can conclude that we found an approximation valid up to $3\%$
for parameter values $c>7$ (We only checked values of $c<25$, but for bigger
$c$\ the approximation will only become better), $-0.5<\rho <0.5$, $T<1$ and
$0.9< K / S_0< 1.1$.

Finally we consider the vanilla call option pricing in LN model combined
with Merton's and Kou's jumps, respectively. Since the jump process is
independent from the approximation we made, we do not investigate the
goodness of our approximation as thoroughly as in the basic LN model
(assuming that, if it is good there it will be good here). Figure \ref%
{f.call} illustrates our analytical results (curves) and the MC
simulations (crosses), as well as the relative errors in the unit of
percent. Each MC simulation runs 300,000,000 times. These results suggest
that the approximation error is typically less than $2\%$. And due to the
fact that whenever the degree of moneyness (the ratio of the strike price $K$
to the initial asset price $S_{0}$) is relatively high, the average bid-ask
spread tends to be relatively high for call options \cite{Pena}, our
analytical results can serve as an easy way to get a quick estimate that is
normally accurate enough for many practical applications.

\section{Conclusion}

\label{s.Conclusion}We presented a method which makes it possible to extend
the propagator for a general SV model to the propagator of that SV model extended with an arbitrary jump process in the asset
price evolution. This procedure, applied to the Heston model, leads to
similar results as those obtained in Ref. \cite{Sepp}, which gives us
confidence in the present treatment. The stationary volatility distribution
of the Heston model, however, does not correspond to the observed lognormal
distribution \cite{Liu, Micciche, Straeten} in the market. The exponential
Vasicek model does have the lognormal distribution as its stationary
distribution. Therefore we used this model for the volatility to illustrate
the method presented in section \ref{s.Results}. For this model no closed
form pricing formulas for the propagator or vanilla option prices exist. We first
derive approximative formulas for the propagator and vanilla option prices for
this model without jumps, using path integral methods. This result was checked with a Monte Carlo simulation, proving a parameter range for
which the approximation is valid. We specified a parameter range for which
our pricing formulas are accurate to within $3\%$.\ They become more accurate
in the limit $\frac{\beta }{\gamma ^{2}}>>1$ where $\beta $\ is the mean
reversion rate and $\gamma $\ is the volatility of the volatility. Finally
we extended this result to the case where the asset price evolution contains
jumps.

\appendix*
\section{Derivation of equations (\ref{algeres}), (\ref{e.U}).} \label{Appendx}

The proof starts by assuming that a solution for $\mathcal{P}_{J}(x_{T},\sigma _{T},T|x_{0},\sigma _{0},0)$ of the form (\ref{algeres}) exists.
Below we show that this assumption indeed leads to a solution, which in turn justifies the assumption.
Since $\int_{-\infty }^{+\infty }\frac{dp}{2\pi }e^{ip(x_{T}-x_{0})}\frac{%
\partial F(\sigma_T, \sigma_0, r, p, T)}{\partial T}$\ equals the right hand side of Eq.(\ref{e.FW})
and\ the derivative operators $\frac{\partial }{\partial x_{T}}$ and $\frac{%
\partial }{\partial \sigma _{T}}$ have no effect on $e^{U(p,T)}$, it follows
that:
\begin{eqnarray}
& &\frac{\partial }{\partial x_{T}}\left[ -\left( r-\frac{1}{2}%
\sigma _{T}^{2}\right) \mathcal{P}_{J}\right]  \notag \\
&&+\frac{1}{2}\frac{\partial ^{2}}{\partial x_{T}^{2}}\left[ \sigma _{T}^{2}%
\mathcal{P}_{J}\right] +\frac{\partial }{\partial \sigma _{T}}\left[
-A(T,\sigma _{T})\mathcal{P}_{J}\right]  \notag \\
&&+\frac{1}{2}\frac{\partial ^{2}}{\partial \sigma _{T}^{2}}\left[
B^{2}(T,\sigma _{T})\mathcal{P}_{J}\right]  \notag \\
&&+\rho \frac{\partial ^{2}}{\partial x_{T}\, \partial \sigma _{T}}\left[
\sigma _{T}B(T,\sigma _{T})\mathcal{P}_{J}\right]  \notag \\
& = & \int_{-\infty }^{+\infty }\frac{dp}{2\pi }e^{ip(x_{T}-x_{0})}\frac{\partial
F(\sigma_T, \sigma_0, r, p, T)}{\partial T}\,e^{U(p,T)}.
\end{eqnarray}

Adding the term $\lambda m^{j}\frac{\partial }{\partial x_{T}}\mathcal{P}_{J}$, which is given by
\begin{eqnarray}
&& \lambda \int_{-\infty }^{+\infty }\frac{dp}{2\pi }ip \, e^{ip(x_{T}-x_{0})}F(%
\sigma _{T},\sigma _{0},r,p,T)\,e^{U(p,T)}  \notag \\
&& \times  \int_{-\infty }^{+\infty }(e^{J}-1)\varpi (J)dJ,
\end{eqnarray}%
as well as the term $\lambda \int_{-\infty }^{+\infty }\left[ \mathcal{P}_{J}(x_{T}-J)-%
\mathcal{P}_{J}(x_{T})\right] \varpi (J)dJ$, which is given by
\begin{eqnarray}
&&\lambda \int_{-\infty }^{+\infty }\frac{dp}{2\pi }%
e^{ip(x_{T}-x_{0})}F(\sigma _{T},\sigma _{0},r,p,T)\,e^{U(p,T)}  \notag
\\
&&\times \int_{-\infty }^{+\infty }\left( e^{-ipJ}-1\right) \varpi (J)dJ,
\end{eqnarray}%
the right hand side of Eq.(\ref{fkarj6}) is expressed as
\begin{eqnarray}
&&\int_{-\infty }^{+\infty }\frac{dp}{2\pi }e^{ip(x_{T}-x_{0})}\frac{%
\partial F(\sigma _{T},\sigma _{0},r,p,T)}{\partial T}\,e^{U(p,T)}
\notag \\
&+&\int_{-\infty }^{+\infty }\frac{dp}{2\pi }e^{ip(x_{T}-x_{0})}F(%
\sigma _{T},\sigma _{0},r,p,T)\,e^{U(p,T)}  \notag \\
&&\times \lambda \int_{-\infty }^{+\infty }\left[
e^{-ipJ}-1+ip(e^{J}-1)\right] \varpi (J)dJ.  \label{rhsn}
\end{eqnarray}%
This, of course should equal the left hand side of Eq.(\ref{fkarj6}), which is given by
\begin{eqnarray}
&&\int_{-\infty }^{+\infty }\frac{dp}{2\pi }e^{ip(x_{T}-x_{0})}\frac{%
\partial F(\sigma _{T},\sigma _{0},r,p,T)}{\partial T}\,e^{U(p,T)}
\notag \\
&+&\int_{-\infty }^{+\infty }\frac{dp}{2\pi }e^{ip(x_{T}-x_{0})}F(%
\sigma _{T},\sigma _{0},r,p,T)\frac{\partial e^{U(p,T)}}{\partial T}.
\label{lhsn}
\end{eqnarray}
Expression (\ref{rhsn}) equals (\ref{lhsn})\ when
\begin{equation}
\frac{\partial U(p,T)}{\partial T}=\lambda \int_{-\infty }^{+\infty }\left[
e^{-ipJ}-1+ip\left( e^{J}-1\right) \right] \varpi (J)dJ,
\end{equation}%
from which the result (\ref{e.U}) for $U(p,T)$ follows.

\begin{center}
\begin{table*}[t]
\caption{Comparison of our approximative analytic pricing result and the MC simulation value for the
LN model. }
\label{T_1}%
\begin{center}
\begin{tabular}{@{}rrrrccc}
\hline
&  &  &  &  &  &  \\
\multicolumn{4}{c}{Parameter values} &  &  &  \\ \cline{1-4}
&  &  &  &  &  & Relative error \\
\, \, $K$ & \quad \quad \, \,$\rho$ & \quad \quad \, \,$\gamma$ & \quad
\quad \, \,$\beta$ & \, \, \, MC value(a)\, & \, \, Approx.(b) \, & \, (b -
a)/a (\%) \\ \hline
90 & -0.5 & 1.2 & 7 & 15.3947 & 15.2533 & -0.9185 \\
&  &  & 8 & 15.2979 & 15.1731 & -0.8166 \\
&  &  & 10 & 15.1630 & 15.0588 & -0.6869 \\
&  & 0.8 & 5 & 15.1079 & 14.9995 & -0.7180 \\
&  &  & 6 & 15.0307 & 14.9337 & -0.6475 \\
&  &  & 7 & 14.9776 & 14.8855 & -0.6151 \\
& 0 & 0.7 & 2 & 15.2486 & 15.1982 & -0.3307 \\
&  &  & 3 & 15.0259 & 15.0024 & -0.1564 \\
&  &  & 4 & 14.9190 & 14.8992 & -0.1328 \\
&  & 0.5 & 1 & 15.2061 & 15.1576 & -0.3187 \\
&  &  & 2 & 14.9030 & 14.8882 & -0.0996 \\
&  &  & 3 & 14.7951 & 14.7865 & -0.0577 \\
& 0.5 & 0.3 & 1 & 14.6051 & 14.5035 & -0.6953 \\
&  &  & 1.5 & 14.5524 & 14.4815 & -0.4872 \\
&  &  & 2 & 14.5354 & 14.4775 & -0.3986 \\
&  & 0.2 & 0.5 & 14.6015 & 14.5398 & -0.4192 \\
&  &  & 0.75 & 14.5609 & 14.5098 & -0.3519 \\
&  &  & 1 & 14.5332 & 14.4981 & -0.2418 \\ \hline
100 & -0.5 & 1.2 & 7 & 9.4541 & 9.2862 & -1.7762 \\
&  &  & 8 & 9.3720 & 9.2197 & -1.6253 \\
&  &  & 10 & 9.2599 & 9.1274 & -1.4310 \\
&  & 0.8 & 5 & 9.1537 & 9.0346 & -1.3006 \\
&  &  & 6 & 9.0950 & 8.9869 & -1.1886 \\
&  &  & 7 & 9.0557 & 8.9534 & -1.1291 \\
& 0 & 0.7 & 2 & 9.5394 & 9.4975 & -0.4395 \\
&  &  & 3 & 9.2906 & 9.2704 & -0.2174 \\
&  &  & 4 & 9.1682 & 9.1513 & -0.1840 \\
&  & 0.5 & 1 & 9.4915 & 9.4493 & -0.4480 \\
&  &  & 2 & 9.1445 & 9.1328 & -0.1285 \\
&  &  & 3 & 9.0235 & 9.0155 & -0.0887 \\
& 0.5 & 0.3 & 1 & 9.0168 & 8.9081 & -1.2051 \\
&  &  & 1.5 & 8.9302 & 8.8522 & -0.8737 \\
&  &  & 2 & 8.8886 & 8.8246 & -0.7195 \\
&  & 0.2 & 0.5 & 8.9655 & 8.9023 & -0.7048 \\
&  &  & 0.75 & 8.9039 & 8.8509 & -0.5955 \\
&  &  & 1 & 8.8628 & 8.8247 & -0.4295 \\ \hline
110 & -0.5 & 1.2 & 7 & 5.2749 & 5.1365 & -2.6237 \\
&  &  & 8 & 5.2209 & 5.0916 & -2.4758 \\
&  &  & 10 & 5.1507 & 5.0335 & -2.2756 \\
&  & 0.8 & 5 & 5.0170 & 4.9219 & -1.8947 \\
&  &  & 6 & 4.9877 & 4.8986 & -1.7862 \\
&  &  & 7 & 4.9709 & 4.8849 & -1.7307 \\
& 0 & 0.7 & 2 & 5.6480 & 5.5989 & -0.8694 \\
&  &  & 3 & 5.3942 & 5.3705 & -0.4394 \\
&  &  & 4 & 5.2684 & 5.2503 & -0.3429 \\
&  & 0.5 & 1 & 5.5967 & 5.5510 & -0.8173 \\
&  &  & 2 & 5.2475 & 5.2343 & -0.2519 \\
&  &  & 3 & 5.1253 & 5.1160 & -0.1821 \\
& 0.5 & 0.3 & 1 & 5.3151 & 5.2095 & -1.9860 \\
&  &  & 1.5 & 5.2048 & 5.1289 & -1.4589 \\
&  &  & 2 & 5.1427 & 5.0812 & -1.1966 \\
&  & 0.2 & 0.5 & 5.2170 & 5.1576 & -1.1380 \\
&  &  & 0.75 & 5.1449 & 5.0954 & -0.9620 \\
&  &  & 1 & 5.0960 & 5.0595 & -0.7163 \\ \hline
&  &  &  &  &  &  \\
&  &  &  &  &  &
\end{tabular}
\\
Other parameter values $S_0 = 100$, $r = 0.015$, $\bar{a} = -1.6$ and $T = 1$ are used here.
\end{center}
\end{table*}
\end{center}

\end{document}